\documentclass[fleqn,11pt]{SelfArx} % Document font size and equations flushed left

\usepackage[english]{babel} % Specify a different language here - english by default

\usepackage{lipsum} % Required to insert dummy text. To be removed otherwise

\definecolor{color1}{RGB}{0,0,90} % Color of the article title and sections
\definecolor{color2}{RGB}{0,20,20} % Color of the boxes behind the abstract and headings

\usepackage{hyperref} % Required for hyperlinks

\hypersetup{
	hidelinks,
	colorlinks,
	breaklinks=true,
	urlcolor=color2,
	citecolor=color1,
	linkcolor=color1,
	bookmarksopen=false,
	pdftitle={Title},
	pdfauthor={Author},
}

\JournalInfo{Accepted for publication in Solar Physics, 2024}
\Archive{DOI 10.1007/s11207-024-02365-0}

\PaperTitle{Is There a Synchronizing Influence of Planets on Solar and Stellar Cyclic Activity?}
\PaperSubTitle{On the Synchronization between Planets and Solar-Stellar Activity}

\Authors{V.N.~\ Obridko\textsuperscript{1}, M.M.~\ Katsova\textsuperscript{2}, D.D.~\ Sokoloff\textsuperscript{1,3,4}*, N.V.~\ Emelianov\textsuperscript{2}}

\affiliation{\textsuperscript{1}\textit{IZMIRAN, 4, Kaluzhskoe Shosse, Troitsk, Moscow, 108840, Russia}}
\affiliation{\textsuperscript{2}\textit{Sternberg Astronomical Institute, Lomonosov Moscow State University, Moscow, 119234, Russia}}
\affiliation{\textsuperscript{3}\textit{Department of Physics, Lomonosov Moscow State University, Moscow, 119991, Russia}}
\affiliation{\textsuperscript{4}\textit{Moscow Center of Fundamental and Applied Mathematics, Moscow, 119991, Russia}}
\affiliation{*\textbf{Corresponding author}: sokoloff.dd@gmail.com}

\newcommand{\keywordname}{Keywords} % Defines the keywords heading name

\Abstract{
This work continues our research of connection between the long-term activity of stars and 
their planets. We analyze new data on the previously considered two dozen solar-type stars with 
identified cycles, adding the results of studying the long-term variability of two more solar-type 
G~stars  and 15 cooler M~dwarfs with planets. If the cyclic activity is determined by a 
strong tidal influence of the planet, then the cycle duration of the star should be 
synchronized with the period of orbital revolution of the planet. We calculate the 
gravitational effect of planets on their parent stars. The results obtained confirm the earlier 
conclusion that  exoplanets do not influence the formation of the stellar cycle. We examine the 
change in the position of the barycenter of the solar system relative to the center of the Sun 
over 420~years. A comparison of these data with the most reliable 120-year SSN (sunspot number) 
series as the index of solar activity  has shown that they are not synchronized.   
}

\begin{document}

\maketitle % Output the title and abstract box

%\tableofcontents % Output the contents section

\thispagestyle{empty} % Removes page numbering from the first page

%----------------------------------------------------------------------------------------
%	ARTICLE CONTENTS
%----------------------------------------------------------------------------------------

\section*{Introduction} % The \section*{} command stops section numbering

\addcontentsline{toc}{section}{Introduction} % Adds this section to the table of contents

Cyclic magnetic activity of the Sun in the form of the Schwabe cycle is one of the most famous 
phenomena of solar physics driven by the solar dynamo. It is believed to be based on 
differential rotation and mirror-asymmetric motions in the solar convection zone. The point, 
however, is that the duration of this cycle (about 11 yrs) is close to the orbital period of 
Jupiter (also about 11 yrs). Solar physicists (e.g. \cite{Hetal19, Retal22}) including the 
members of our team usually consider this fact to be just a coincidence, however some are 
inclined to believe that Jupiter does contribute to the physical processes responsible for the 
cycle (e.g. \cite{SB22, Ketal23} and references therein). Indeed, it is quite difficult to 
prove that one physical phenomenon does not influence another, and there is no need to argue 
that planets cannot anyhow affect the solar activity in principle.

The progress in the present-day astronomy makes it possible to have a broader look at this old 
controversy. Indeed, the cyclic activity is known to exist at dozens of stars more or less 
similar to the Sun, and some of them have exoplanetary systems confirmed by observations. This 
fact allows us to enlarge substantially the observational basis of the discussion. 
\cite{Oetal22}
carried out a corresponding analysis of observational data and found no evidence that stellar 
cycles could be considered a manifestation of planetary influence on the interior of the star. 
That article was the first step in this direction, but it did not cover all aspects of the 
problem. It still deserves a more systematic study, which is even more motivated taking into 
account a rapid progress in this field.

We appreciate that the planetary explanation of the solar cycle looks quite implausible in 
context of contemporary solar physics and one could say that the burden of proof is on 
proponents of the theory. Taken however into account that the topic under discussion is 
interesting and important for quite a broad audience, we believe that the arguments based on 
direct astronomical observations rather than on theoretical ideas could be important here. 

This is the motivation of the present article. In particular, we analyze the latest data on 
activity of the previously considered solar-type  G and K stars and complement them 
with data on the long-term variability of 15 cooler M dwarfs with planets. We carry out a 
detailed analysis of the possible tidal influence of a planet on the parent star, 
including data on the barycenter position of the planetary system. We analyze the barycenter 
position of the solar system over 420 years and compare it with available sunspot data.

\section{Observations of Stellar Activity Versus Planetary Data}

%\addcontentsline{toc}{section}{Observations of Stellar Activity Versus Planetary Data}

\subsection{Updating the Magnetic Cycle Data}

%\addcontentsline{toc}{subsection}{Updating the Magnetic Cycle Data}

First, we have updated the list of stars with known magnetic activity and the exoplanetary 
data used in our previous work \cite{Oetal22} based on the stellar activity data from 
\cite{Betal95}. 
This dataset is complemented with the latest data from \cite{Betal22} to add stars 
that were not considered by \cite{Betal95}. Obviously, it is important to demonstrate that 
using \cite{Betal22} as the preferred source of information, we arrive at the same conclusions. 
A comparison is made (Table 1) to show that for stars with known exoplanetary systems, the 
changes in the magnetic activity revealed by \cite{Betal22} do not affect the conclusion about 
the lack of connection between the planets and stellar activity.

More precisely, \cite{Betal22} slightly modified the cycle length estimates for particular 
stars; however, these changes are comparable with cycle-to-cycle variations in the length of 
the solar cycle. For HD 190406, the data accumulated allow us to draw a fairly definite 
conclusion about the type of stellar activity. For HD 166620, it is possible to show that the 
star shows a behavior similar to that of the Sun during the Maunder minimum. Both conclusions 
do not help us relate the phenomenon of the stellar cycle with planetary influence. 

When compiling Table \ref{T1}, we have ignored the suggestion of \cite{Betal22} to separate 
naked-eye estimates of cycle length from cases where the data allow periodogram analysis. 
Obviously, the amount of data accumulated has enabled the use of more sophisticated processing 
techniques. However, historically, the discussion of planetary influences on the solar cycle 
began in the 19th century, based on data comparable in quality to modern data on stellar 
activity. 

\begin{table}[!t]
\caption{
Comparison of the activity data for stars with planetary systems from \cite{Betal95} and 
\cite{Betal22}. E means that the authors define the cycle as "Excellent", i.e., the most reliable 
\cite{Betal95}. \cite{Betal22} use this definition by default. MM indicates that we are 
presumably dealing with an event of the type of the Maunder minimum \cite{Letal22}. 
Here $P_{\rm cyc}$ is the length of the activity cycle.
}
\begin{center}
\begin{tabular}{|l|l| l|l|l|}
\hline
 &\multicolumn{2}{l|}{Baliunas at al. 1995}&\multicolumn{2}{l|}{Baum et al. 2022} \cr
 \hline
 Star & $P_{\rm cyc}$, yrs & Activity type & $P_{\rm cyc}$, yrs & Activity type \cr
 \hline
 HD 4628 & 8.4 & E & 10.0 &  E \cr
 \hline
 HD 10476 & 9.6 & E & 10.3 &  E\cr
 \hline
HD 26925 & 10.1 & E & 9.9 &  E\cr
 \hline
HD  166620 & 15.8 & E  & 17.0 & E,  MM \cr
\hline
HD 190406 & 3.6 + 16.9 & fair+good & 17.2 &  E \cr
\hline
HD 152391 & 10.9  & E & 9.1 &  E \cr
\hline
HD 219834 &  10.0 & E & 9.4 &  E \cr
\hline  
\end{tabular}
\end{center}
\label{T1}
\end{table}

\subsection{Exoplanets and Activity of M Dwarfs}

In \cite{Oetal22}, we considered G and K solar-type stars with known cycles. The present 
analysis  includes data on the magnetic activity of M dwarfs \cite{IS23} hosting planetary 
systems  (Table~\ref{T2}). Magnetic activity of M dwarfs seems to be a natural extension of 
the topic under discussion. Note that the data presented in Tables \ref{T2} and \ref{T3} are 
the result of long-term photometric monitoring, while the data considered above were obtained 
from the analysis of chromospheric activity. The interpretation of  stellar activity in terms 
of the  dynamo theory suggests that the physical processes responsible for magnetic activity 
on stars occur in the interior of the star much lower than the photosphere, to say nothing of 
the chromosphere. However, it seems plausible to suggest that photospheric and chromospheric 
data represent the stellar activity. The current body of observational data appears to be 
insufficient to decide how important this difference is overall. In any case, we do not see 
any pronounced difference that is important in the context of the planetary hypothesis.

%\begin{tiny}
\begin{table}[!p]
\caption{M dwarfs with planets and the type of activity known from \cite{IS23}.  $P_{\rm rot}$ 
is the rotation period, $P_{\rm cyc}$ is the cycle length, $M_p$ is the mass of the planet (
stands in the rows for planet), $V_p$ (given in bold; see text for the definition) is the 
ratio of the planetary influence on the star to the influence of Jupiter on the Sun (stands in 
the stellar row), $R_p$ is the radius  of the planet, $P_{\rm orb}$ is the orbital period, and 
$R_{\rm orb}$ is the orbital radius in astronomical units (AU). The activity (stands in the 
stellar row)  is denoted as follows: E stands for Excellent, G for Good, F for Flat, L for 
Long, P for poor, and V for Var (based on the probability of a false alarm ).  Me is the mass 
of the Earth, Re is the radius of the Earth, Mj is the mass of Jupiter, Rj is the radius of 
Jupiter,  and * means that the orbital period is estimated by the Kepler's law.
}
\begin{center}
\begin{tabular}{|l|l|l|l|l|l|l|}
\hline
Name & $P_{\rm rot}$ & $P_{\rm cyc}$ & $V_p$ or $M_p$ & $R_p$ & $P_{\rm orb}$ & $R_{\rm orb}$ \cr
&d &yrs&&&& AU\cr
\hline
GJ 581* (M3 V) &141.6&3.8  & {\boldmath $9186$}&&& \cr
GJ 581 b&&&15.9 Me& 0.366 Rj& 5.4 d& 0.040  \cr
GJ 581 c &&&5.5 Me & 2.21 Re & 12.7 d& 0.07  \cr
GJ 581 e &&&1.7 Me & 1.17 Re & 3.1 d& 0.028  \cr
\hline
GJ 628 (M3 V)&79.6 & 3 & {\boldmath$ 961$}&&&G \cr
Wolf 1061 &&&&&& \cr
Wolf 1061 b &&&1.91 Me& 1.21 Re &4.9 d& 0.03  \cr
Wolf 1061 c &&& 3.41 Me & 1.66 Re & 17.9 d& 0.08  \cr
Wolf 1061 d &&& 7.7 Me & 0.24 Rj & 217.2 d & 0.04  \cr
\hline
GJ 849* (M3.5 V) & 41.4 & 3.7   & {\bf 3.37}& && P \cr
GJ 849 b &&& 0.9 Mj & 1.24 Rj& 1914 d & 2.35 \cr
GJ 849 c &&& 0.702 Mj & 1.25 Rj & 7049 d & 4.9  \cr
\hline
GJ 896A* (M3.5 V)& 15.58 & 10.1   & {\bf 638} & && L \cr
GJ 896A b & && 2.26 Mj& 1.19 Rj& 284.4 d& 0.63\cr
\hline
GJ 317* (M3.5 V)& 57.5 & 2.3  & {\bf 109}& &&L \cr
GJ 317 b &&& 1.7528 Mj & 1.151 Rj & 695.66 d & 1.151  \cr
GJ 317 c &&& 1.644 Mj & 1.2 Rj & 18.5 yrs & 5.23 \cr
\hline
GJ 273 (M3.5 V)&115.9 & 5.8 & {\bf 13} &&&G \cr
GJ 273 b &&&0.0069 Mj & 1.51 Re &18.64 d &0.09  \cr
GJ 273 c &&& 1.18 Me &1.06 Me & 4.7 d& 0.036  \cr
GJ 273 d &&& 0.0345 Mj & ? & 413.9 d& 0.712  \cr
GJ 273 e &&& 0.0297 Mj & ? & 542 d & 0.849  \cr
\hline
GJ 447 (M4 V) & 175.9  & 5.3   & {\boldmath $156$} &&& P \cr
Ross 128  &&&&&&\cr
Ross 128 b &&& 1.11 Me & 1.4 Re & 9.9 d& 0.049  \cr
\hline
GJ 54.1 (M4 V)& 2.78  & 10.4  & {\boldmath $1420$}& && G \cr
YZ Cet &&&&&& \cr
YZ Cet b &&& 0.913 Me & 0.7 Re & 2 d & 0.016 \cr
YZ Ce tc &&& 1.05 Me & 1.14 Re & 3.1 d & 0.02  \cr
YZ Cet d &&& 1.03 Me & 1.09 Re & 4.7 d& 0.028  \cr
\hline
GJ 551 (M4.5 V) & 85.1 & 5   & {\boldmath $7 \times 10^5$}&&&L  \cr
Prox Cen &&&&&& \cr
Prox Cen b &&& 1.03 Me & 1.07 Re & 11.2 d & 0.004  \cr
Prox Cen d &&& ? & 0.0008 Rj & 5.167 d & 0.028  \cr
\hline
GJ 406 (M6 V) & 7.4 & 13   & {\boldmath $1 \times 10^5$}&&& F \cr
Wolf 359 &&&&&&\cr
Wolf 359 b &&&  0.138 Mj && 2938 d & 1.845 \cr
Wolf 359 c &&&  0.012  Mj& & 2.687 d & 0.018  \cr
\hline
\end{tabular}
\end{center}
\label{T2}
\end{table}
%\end{tiny}

\subsection{Additional Data Concerning G Stars}

Further expansion of the observational data base of our study is possible by including 
photometric data on stellar activity sufficient to identify the stellar activity cycle. 
\cite{Letal16} provided data on the magnetic activity of two G stars with confirmed planets, 
which are a useful addition to the data listed in Table~\ref{T1} and the data used by 
\cite{Oetal22} (Table \ref{T4}).

%\begin{tiny}
\begin{table}[!t]
\caption{G stars with planets and the type of activity known from \cite{Letal16} (based on  
photometric data). Designations are the same as in Table~\ref{T2}. For comparison, we include 
the case of solar activity and Jupiter in the table.
}
\begin{center}
\begin{tabular}{|l|l|l|l|l|l|l|}
\hline
Name & $P_{\rm rot}$ & $P_{\rm cyc}$ & $V_p$ or $M_p$ & $R_p$ & $P_{\rm orb}$ & $R_{\rm orb}$ \cr
&d &yrs&&&& AU\cr
\hline
\hline
HD 63433 (G5 V) & 6.46 & 5   & {\boldmath $5029$}&&&P\&L  \cr
V377 Gem &&&&&& \cr
HD 63433b &&& 5.11 Me & 2.112 Re & 7.1 d & 0.0714  \cr
HD 63433c &&& 6.9 Me & 0.225 Rj  & 20.5 d & 1.1448  \cr
HD 63433d &&& 1.25 Me & 1.073 Re  & 4.2 d & 0.0503  \cr
\hline
HD 70573 (G6 V) & 3.31 & 6.9   & {\boldmath $127$}&&& G \cr
V478 Hya &&&&&&\cr
HD  70573 b &&&  6.1 Mj&1.14 Rj& 2.3 yrs & 1.76 \cr
\hline
The Sun & 25 d & 11 yrs & {\bf 1} & &&  E \cr
Jupiter & && 1 Mj &  1 Rj & 11.86 yrs &  5.2 \cr 
\hline
\end{tabular}
\end{center}
\label{T3}
\end{table}
%\end{tiny}

\subsection{Stars with Cyclic Activity without Planets}

The observational base of our research can also be expanded by including stars with an 
established type of magnetic activity, for which the search for a planetary system was not 
successful (Table~\ref{T4}). We understand that the available planet detection method may fail 
to detect an  existing planet. However, we believe that such option should be taken into 
account. Indeed, if there are stars without planets but with a pronounced magnetic cycle, it 
becomes problematic to insist that the cycle as a physical phenomenon is associated with 
planets. In Table \ref{T4} we see at least one example  (GJ 285) that confirms this point of 
view. Let us note here that in modern scientific literature it is not easy to identify cases 
of unsuccessful attempts to find exoplanets near a star. Perhaps by paying more attention to 
this search, the table could be expanded.

\begin{table}[!t]
\caption{Red dwarf stars with the type of magnetic activity known from \cite{IS23} without a 
planetary system revealed by available observations. Designations are the same as in Table 1.
}
\begin{center}
\begin{tabular}{|l|l|l|l|l|}
\hline
Name & $P_{\rm rot}$ & $P_c$& Activity type \cr
\hline
GJ 358 (M4 V)& 25.2 d & 4.7  yrs&  P \cr 
\hline
GJ 729 (M3.5 V) & 2.9 d & 3.5  yrs & F \cr
\hline
LP 816-00 (M4 V)& 86.3 d & 2.0  yrs & F \cr
\hline
GJ 285 = YZ CMi (M4 V) & 2.78 d & 10.4  yrs & G \cr
\hline 
GJ 234=Ross 614 (M4.5 V)& 1.58 d & 5.9  yrs & F \cr
\hline 
\end{tabular}
\end{center}
\label{T4}
\end{table}

\subsection{Summarizing New Observations}

By summing up the data mentioned above (17 star systems) and the data used in \cite{Oetal22} 
(15 star systems and the  solar system), we actually double the number of cases under 
consideration, although the resulting database becomes less homogeneous than in the latter 
mentioned article.

A preliminary analysis of the complete list reveals the following points. First of all, not 
all of the listed stars (33 cases) display a pronounced magnetic cycle, designated in the 
tables as G and E (10 cases). The above proportion can hardly be considered a correct estimate 
of the share of stars with pronounced cycles among all candidate stars with periodic activity. 
However, we believe that this relationship can be used as an approximation. Apparently, there 
is something else to take into account account. i.e. the rotation of the star. This condition 
is entirely consistent with the dynamo interpretation of activity cycles and seems problematic 
in terms of the planetary hypothesis.

The parameters of stars and planets are taken from the Exoplanet Search Catalogues, a link to 
which is provided in the Data Availability declaration.

The tables do not provide clear support for the planetary hypothesis. Perhaps we can take a 
closer look at the following cases. 

Case HD 176051AB (G0 V), i.e. Jupiter at the orbit of Mars, seems most promising. According to 
the planetary hypothesis, one would expect a cycle length of about 3 years, while the actual 
cycle is about 10 years, and observers do not consider these numbers to be particularly 
reliable. 

Case GJ 628 (M3 V), i.e. the Earth inside the orbit of Mercury, and case GJ 896A (M3.5~V),
i.e. Jupiter at the orbit of Venus, give an activity cycle several times longer than the 
orbital period. Case GJ 273 (M3.5~V) involves many planets and opens the way to various 
speculations. However, it does not allow us to simply identify the duration of the activity 
cycle with the revolution period. In two out of three cases, we have the periodicity of type 
G. 

As regards the interpretation in terms of the planetary hypothesis, we estimate the  tidal 
effect $V_P$ caused by the gravity of planetary systems and compare it with the effect arising 
from the influence of Jupiter on the Sun (see Tables \ref{T2} and \ref{T3}). The quantity $V_P$
 is estimated as the ratio $M r^2 (3 \cos^2 \phi -1)/R^3$ calculated for a given planetary 
system and is related to the corresponding quantity in the solar system. The scaling can be 
found in various textbooks, see e.g. \cite{MD99}. Here $r$ is the stellar radius, $M$ is the 
planetary mass, $R$ is the orbital radius, and $\phi$ is the angle between the orbital plane 
and the stellar equator.  Note that the tidal potential on the surface does not depend on the 
mass of the star. If a star has several planets we give the estimate for the planet with the 
maximum gravitation potential.

Analysis of the data obtained and their comparison with the data provided in \cite{Oetal22} 
confirms that the gravitational effect $V_p$ varies in a wide range and there is no clear 
connection between $V_p$ and the properties of activity cycles of the parent stars. 
The only more or less clear message is that Jupiter's influence on the hydrodynamics of 
its parent star (the Sun) is, perhaps, not the largest on the list.

To summarize the above, we can say that we have made every effort to find in the available 
observational data any confirmation of the planetary hypothesis, but failed. Therefore, 
we have to consider the fact that the solar magnetic activity cycle is close to the 
orbital period of Jupiter as a simple coincidence, at least until some radically new data 
arrive.

\section{Planets and Details of the Solar Cycle History}

In this section,  we will compare the characteristics of solar activity with the total 
influence of the planets determined not only by Jupiter. Note an important difference between 
the solar cycle and the barycenter motion: the latter is much more stable than the solar 
cycle. Indeed, the length of the solar cycle has varied by a year or two (sometimes more) 
during the history of instrumental  observations of the Sun (about four centuries). For the 
same period, the cycle amplitude has varied much more substantially: at least by an order of 
magnitude from the Maunder minimum at the middle of  the 17th and the beginning of the 18th 
century to very high cycles in the second half of the 20th  century. In principle, this 
approach provides much more opportunities for relating some features in solar cycle variations 
with planetary effects. The problem is to isolate a specific feature, associate it with 
another feature in the planetary motion, and propose a physical mechanism for this connection. 
We are not talking here about any specific hypothesis with a developed physical mechanism. 
There are many assumptions that more or less convincingly connect the dynamics of the solar 
cycle with the planetary motion 
(e.g. \cite{P65, z97, CP18, O20}). In particular, we agree with the idea proposed in the 
works cited above that the position of the barycenter of the solar system inside or outside 
the Sun is decisive for the shape of the cycle.

\begin{figure}
        \centering 
        \includegraphics[width=0.85\textwidth]{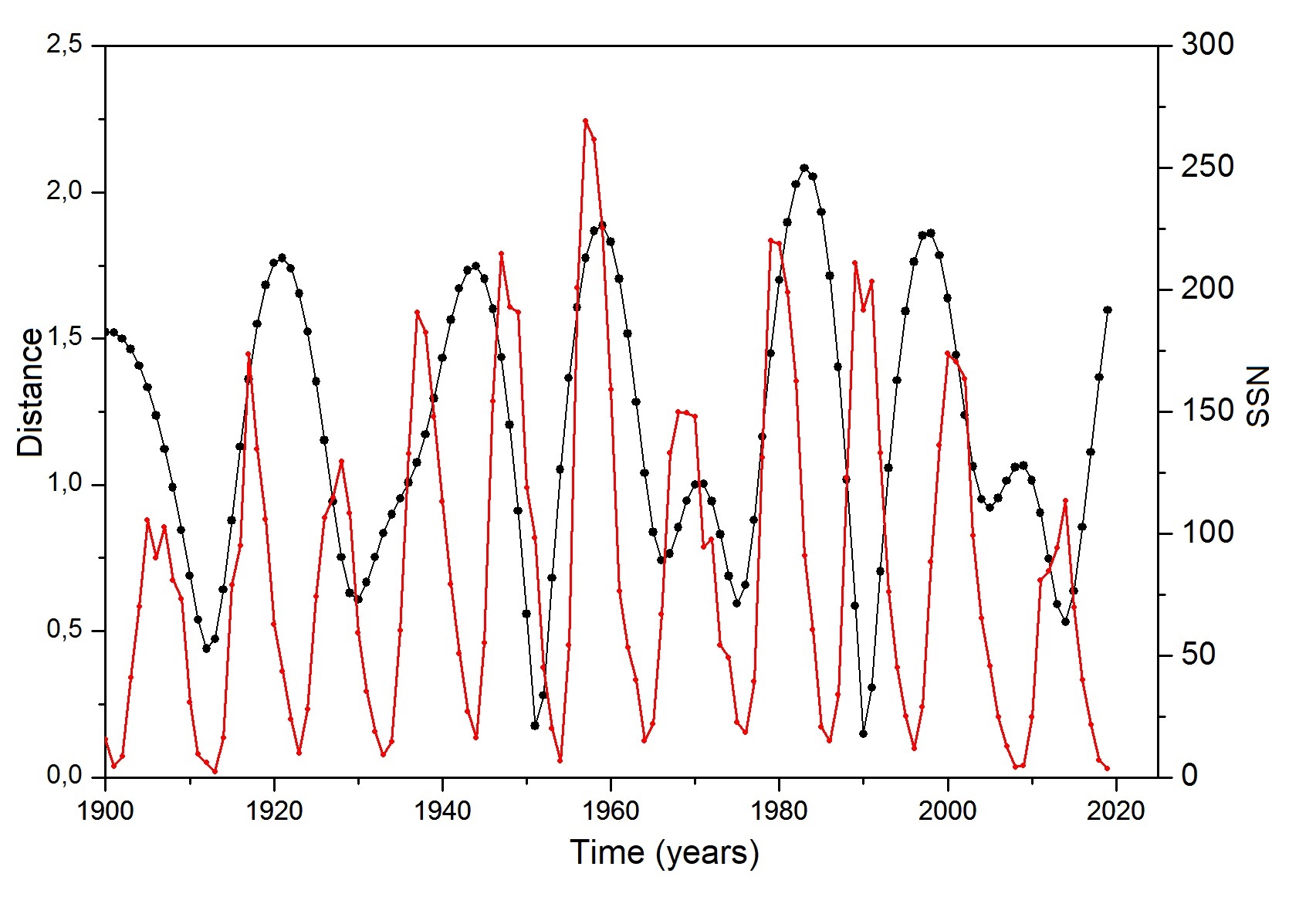}
        \caption{Evolution of the barycenter of the solar system (the distance from the solar
		center, black) and the sunspot number (SSN, red) from 1900.}
        \label{F1}
    \end{figure}

Of course, if the orbital period of Jupiter is close to the duration of the solar cycle, the 
barycenter motion and, say, the sunspot number record do demonstrate some similar features. 
Our point, however, is that both quantities display something more complicated than just a 
harmonic oscillation. Then, it seems reasonable to compare the phase behavior of both tracers 
(namely, the distance between the solar center and barycenter of the the solar system  and the 
sunspot number (SSN), see Figure~\ref{F1}). The barycenter distance is calculated based on 
\cite{Fetal14}. We do not see here any pronounced phase relationship between both signals. 

\begin{figure}
        \centering
        \includegraphics[width=0.45\textwidth]{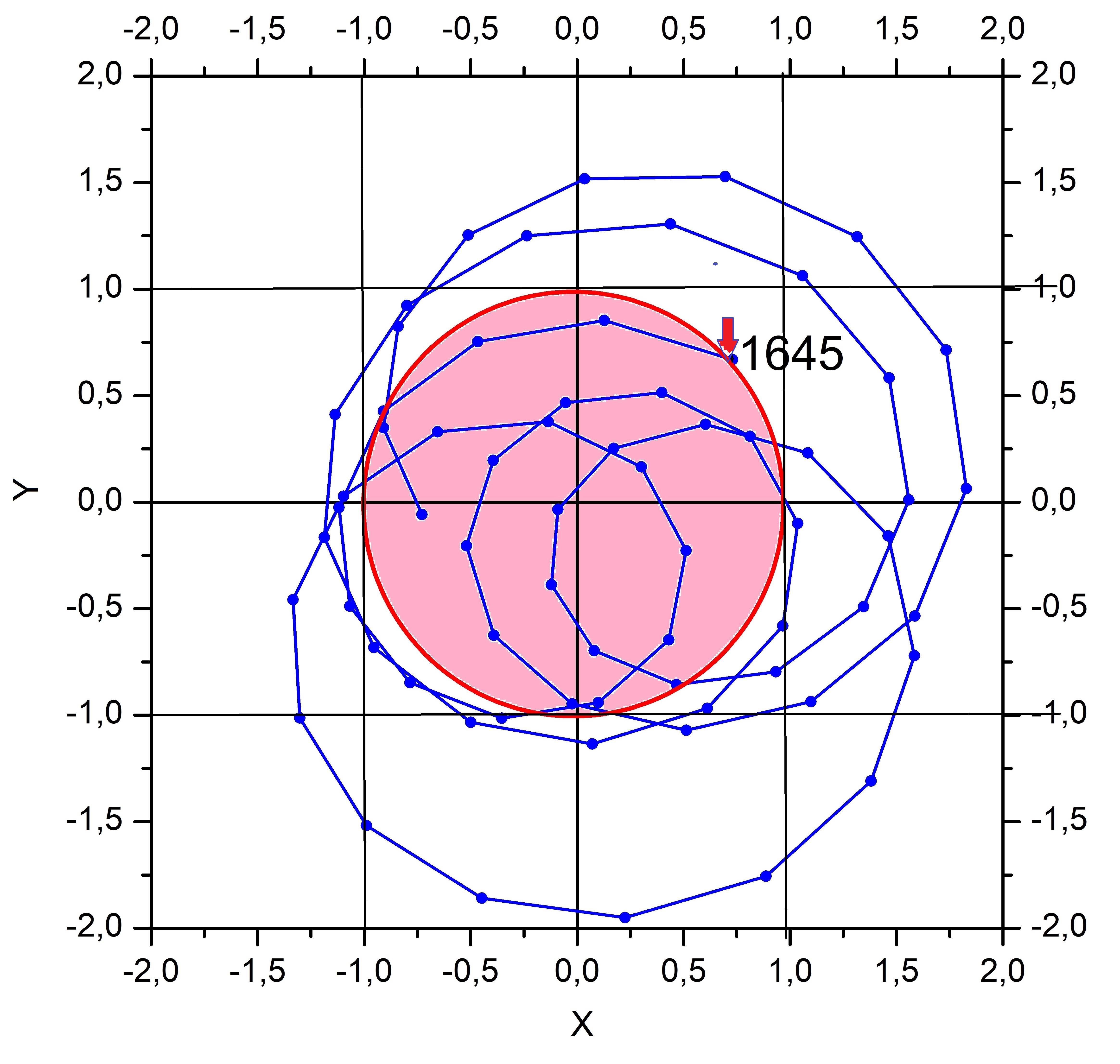}
         \includegraphics[width=0.45\textwidth]{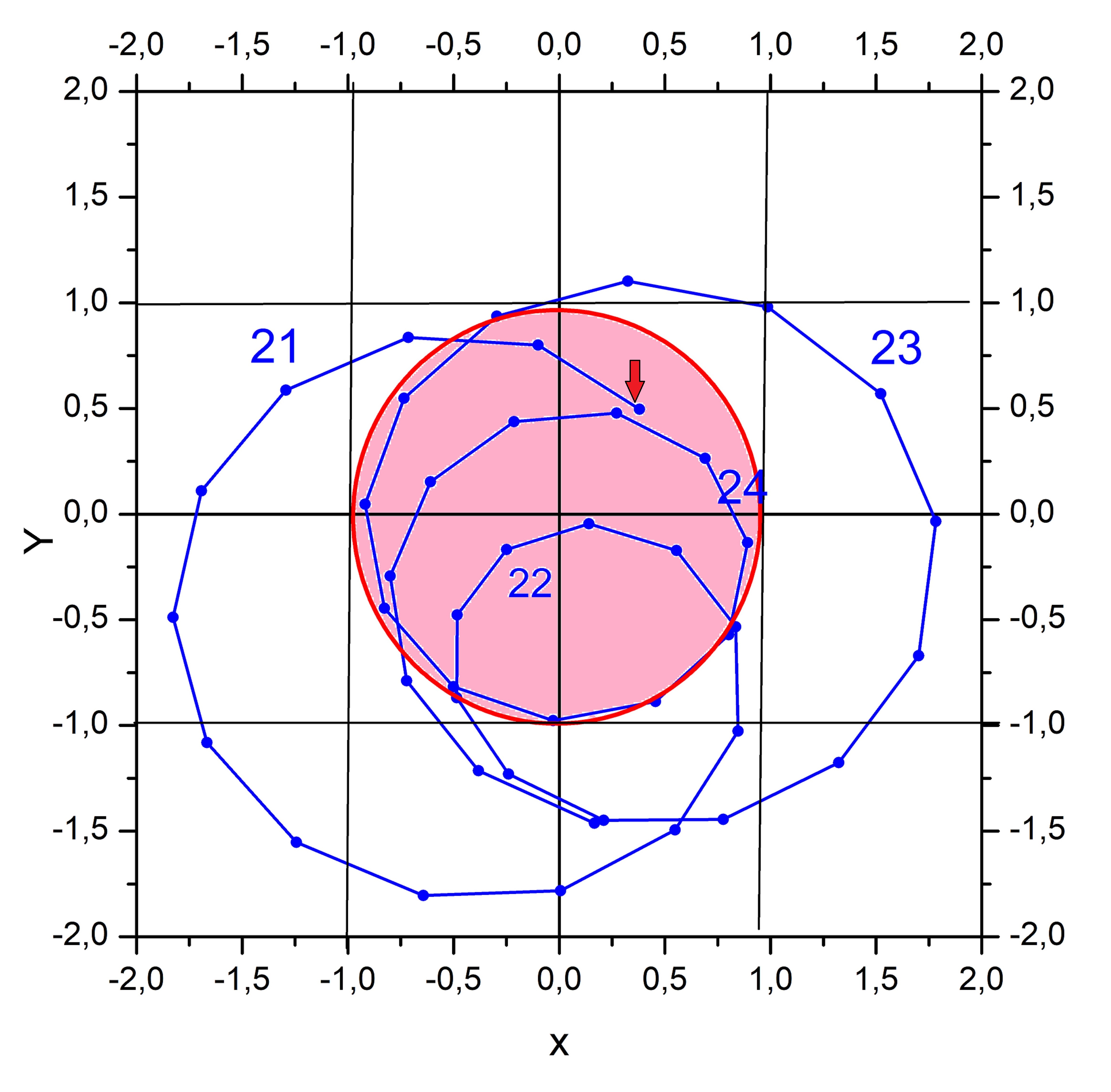}
        \caption{Motion of the solar system barycenter in the solar equatorial plane during 
the Maunder minimum (left, the starting point, 1645, is marked) and several recent 
solar cycles (right, cycles numbers are indicated). The solar interior is shown as a 
red circle.  The $x$ axis is directed from the Sun to the Earth. }
        \label{F2}
    \end{figure}

More precisely, the idea could be formulated as an assumption (e.g. \cite{O20}) that during 
such outstanding events in the history of solar activity as the Maunder minimum, the 
barycenter was located in some special way (for example, inside the Sun). In order to verify 
this assumption, we present in Figure~\ref{F2} the evolution of the solar system barycenter in 
the solar equatorial plane for two different epochs in the evolution of solar activity -- the  
Maunder minimum (left panel) and the contemporary epoch (right panel). The equatorial 
cross-section of the Sun is shown as a red circle. 

Strictly speaking, the time boundaries of the Maunder minimum are quite arbitrary. In Figure~
\ref{F2}a, we choose 1645-1710 as the conditional time boundaries. The starting point is shown 
with a vertical red arrow, the points mark every year. In Figure~\ref{F2}b, the points 
correspond to the period of the Solar Cycles 21\,--\,24 (1976\,--\,2019); the first point is 
also shown with a vertical red arrow.

Of course, the trajectories differ; however, it seems difficult to choose which one is more 
favorable to drive the solar cycle, since the sunspot statistics for both epochs is very 
different. A comparison of Figures~\ref{F2}a and \ref{F2}b shows that there is no reason to 
assert that Maunder-type grand minima arise as the barycenter moves away from the Sun.

Again, we agree that barycenter distance as an important participant in the excitation of the 
solar cycle looks quite implausible in context of celestial mechanics (corresponding 
argumentation can be found in e.g. \cite{S06}). However it might be helpful to support this 
theoretical argumentation by the fact that the behaviour of the barycenter anticipated by 
supporters of the theory does not follow from the celestial mechanics equations. 

\section{Conclusion and Discussion}

To summarize, we find no reason to consider the planetary effect as the main driving force 
of the solar magnetic cycle. In this sense, we confirm the earlier conclusion made in 
\cite{Oetal22} and expand the observational base for this conclusion. In particular, in 
addition to stars with a magnetic periodicity different from the planetary orbital period, we 
present an example of a star with a magnetic cycle, which, however, does not have a planetary 
system (at least the search for exoplanets did not find any).

The main point is that the planetary hypothesis does not propose any clear mechanism 
independent of the stellar dynamo that would convert the mechanical force of the planetary 
motion into magnetic field variations.

We do not deny that planets may somehow affect the stellar activity cycle driven by motions 
unrelated directly  to planetary effects. Again, we support our earlier conclusions 
\cite{Oetal22} and note a shift in part of some followers of the planetary hypothesis 
in this more fruitful (as we believe) direction (e.g. \cite{Setal21, Ketal23}). 
Indeed, having 
oscillations with two close periods of about 11 years one can easily obtain beatings with a 
period of about 100 years (e.g. \cite{Setal21}), which could be associated with well-known 
Gleissberg cycle (e.g. \cite{H10}). However, the available bulk of observational data do not 
provide unambiguous confirmation of this idea. On the one hand, observers propose candidate 
stellar events similar to the Maunder minimum (e.g. \cite{Letal22}) where long-term beatings 
are unlikely to be associated with the relationship between the cycle length and the orbital 
period. On the other, the behavior of the solar activity reconstructed for 10.000 years based 
on isotopic data (see for review \cite{U23}) demonstrates a number of the Maunder-type grand 
minima, whose distribution on the time axis looks random rather than periodic as one could 
expect after a straightforward application of the idea of beatings. We certainly do not deny 
the existence of long cycles such as the Suess cycle. It is just that contemporary data are 
not enough to confidently detect them and establish a quantitative range of periods. At present
-day level of knowledge, it seems reasonable to avoid strong statements and consider the 
planetary effect to be a possible factor in the physics underlying the behavior of the long-
term stellar cycle.

One more point to mention here is that the data concerning  exoplanetary systems is biased by 
various selection effects. Further prog\-ress in exoplanetary studies hopefully should allow 
to take selection effects into account and may, in particular, to support (or reject) our 
conclusions.

Of course, the selective effect cannot be completely excluded. However, the observation 
conditions are such that it is selection that leads to the discovery of the most massive and (
or) closest to the star planets. Additionally, these planets should make the most significant 
contribution to the potential. 

Of course, we are aware that although qualitatively both the tidal effect and
the detection probability increase the mass of the planet ($M$) and decrease 
with the orbital radius (R), the
quantitative dependence varies. The tidal effect scales as $M/R^3$
while, e.g., the radial velocity amplitude caused by a planet scales
as $M^{3/2}/R^{1/2}$, the the birghtness of the planet
for direct detection scales as
$M^{2/3}/R^2$ etc. So for a given sensitivity threshold, the allowed
area in the $M$-$R$ plane will be different for each of these.

Our selection includes exoplanets in a very wide range of masses, sizes and distances. 
It is difficult to imagine that such a set arose as a result of selection.

\section*{Acknowledgements}
We thank the anonymous reviewer for very useful comments.

%\section*{Fundinginformation}
DDS thanks the financial support of the Ministry of
Education and Science of the Russian Federation as part of the program
of the Moscow Center for Fundamental and Applied Mathematics under the
agreement No 075-15-2022-284.

%\begin{conflict}
    The authors declare that they have no conflict of interests.

\section*{Data Availability}
The data on sunspots were taken from WDC-SILSO, Royal Observatory of Belgium, Brussels \\
\url{https://sidc.be/SILSO/datafiles}. \\
Search for exoplanets around stars was carried out in databases of  NASA Exoplanet Archive \\
\url{https://exoplanets.nasa.gov/discovery/exoplanet-catalog/} \\
and  Extrasolar planet catalogues \url{http://exoplanet.eu/catalog/}. \\
We used stellar activity data from \cite{Betal95, Letal16, Betal22, IS23}. 
In this research, we used the SIMBAD database, operated at CDS, Strasbourg, France,
and NASA’s Astrophysics Data System Bibliographic Services.

%----------------------------------------------------------------------------------------
%	REFERENCE LIST
%----------------------------------------------------------------------------------------

\phantomsection
\bibliographystyle{unsrt}
%\bibliography{sample}
%\bibliographystyle{spr-mp-sola}
\bibliography{bibliography_ex}

%----------------------------------------------------------------------------------------

\end{document}